\documentclass[11pt]{article}
\usepackage{moriond,epsfig,mathrsfs,wrapfig}

\def\PRL{{\em Phys. Rev. Lett.}}
\def\PRD{{\em Phys. Rev.} {\bf D}}

\def\be{\begin{equation}}
\def\ee{\end{equation}}
\def\bea{\begin{eqnarray}}
\def\eea{\end{eqnarray}}

\def\ptmin{p_{T{\rm min}}}
\begin{document}
\hspace*{\fill}\parbox[t]{4cm}{
IPPP/03/17\\
DCPT/03/34\\ 10th April 2003}
\vspace*{3.5cm}
\title{A Z-MONITOR TO CALIBRATE HIGGS PRODUCTION VIA VECTOR BOSON FUSION WITH
RAPIDITY GAPS AT THE LHC}

\author{ P.H. WILLIAMS }

\address{Institute for Particle Physics Phenomenology, University of Durham, South Road,\\
Durham DH1 3LE, England}

\maketitle\abstracts{ We study central $Z$-boson production accompanied by
  rapidity gaps on either side as a way to gauge Higgs weak boson fusion
  production at the LHC. We analyse the backgrounds for the $b\bar{b}$-decay
  mode and show that these can be substantially reduced.  Special attention
  is paid to the evaluation of the gap survival factor, which is the major
  source of theoretical uncertainty in the rate of $H$ and $Z$ central
  production events with rapidity gaps.}

\section{Introduction}
Hunting the Higgs boson(s) is now the highest priority of the international
high-energy physics programme. To ascertain whether a Higgs signal can be
seen, it is crucial to show that the background does not overwhelm the
signal. For instance, the major difficulty in observing inclusive production
of the Higgs in the preferred mass range around 115~GeV via the dominant
$H\rightarrow b\bar{b}$ mode is the huge $b\bar{b}$ QCD background. An
attractive possibility to reduce the background is to study the central
production of the Higgs in events with a large rapidity gap on either side.
An obvious advantage of the rapidity gap approach is the clean experimental
signature -- hadron free zones between the remnants of the incoming protons
and the Higgs decay products.\par The cross section is large enough in the
semi-inclusive case when the protons dissociate,
\begin{equation}
\label{eqn:semi_inc} 
p p\rightarrow X + H + Y 
\end{equation}
where the plus sign denotes a large rapidity gap. A significant contribution
to process~(\ref{eqn:semi_inc}) comes from Higgs production via $WW/ZZ$
fusion, i.e. $qq\to qqH$. Since this process is mediated by colourless
$t$-channel $W/Z$ exchanges there is no corresponding gluon bremsstrahlung in
the central region, and thus Sudakov suppression of the rapidity gaps does
not occur. Another characteristic feature of the vector boson fusion Higgs
production process is that it is accompanied by energetic quark jets in the
forward and backward directions.\par The most delicate issue in calculating
the cross section for processes with rapidity gaps concerns the soft survival
factor $\hat{S}^2$. This factor has been calculated in a number of models for
various rapidity gap processes, see for example~\cite{18,19,21,22}. Although
there is reasonable agreement between these model expectations, it is always
difficult to guarantee the precision of predictions which rely on soft
physics.\par Fortunately, calculations of $\hat{S}^2$ can be checked
experimentally by computing the event rate for a suitable calibrating
reaction and comparing with the observed rate. As shown in~\cite{9,20} the
appropriate monitoring process for the double-diffractive mechanism is
central dijet production with a rapidity gap on either side. To date, such a
check has been the prediction of diffractive dijet production at the Tevatron
in terms of the diffractive structure functions measured at HERA~\cite{23}.
The evaluation of the survival factor $\hat{S}^2$ based on the formalism
of~\cite{18,19} appears to be in remarkable agreement with the CDF data. We
expect that future measurements in Run II of the Tevatron will provide us
with further detailed information on $\hat{S}^2$.\par As was pointed out
in~\cite{26,27}, the survival factor for the gaps surrounding $WW\rightarrow
H$ fusion can be monitored experimentally by observing the closely related
central production of a $Z$ boson with the same rapidity gap and jet
configuration.\par We develop these ideas further by considering the decays
of both (light) Higgs and $Z$ bosons into $b \bar b$ pairs, the dominant
decay channel of the former.  In each case we require two forward energetic
jets, and rapidity gaps on either side of the centrally produced decay
products. Both $H$ (Fig~\ref{fig:h}) and $Z$ (Fig~\ref{fig:zew}) can be
produced by electroweak vector boson fusion, for which gaps are `natural',
but the $Z$ can also be produced via $\mathscr{O}(\alpha_{S}^{2}\alpha_W)$
QCD processes (Fig~\ref{fig:zqcd}), with both quarks and gluons exchanged in
the $t$-channel. Finally, there is a large continuum
$\mathscr{O}(\alpha_{S}^{4})$ $b \bar b$ background (Fig~\ref{fig:hbk}).
\begin{figure}[ht]
  \begin{minipage}[b]{0.25\linewidth}
    \begin{center}
      \scalebox{0.6}{\includegraphics{wwfus_make.epsi}}
      \caption{Higgs production via electroweak vector boson fusion.}\label{fig:h}
    \end{center}
  \end{minipage}
  \hspace{5mm}
  \begin{minipage}[b]{0.7\linewidth}
    \begin{center}
      \scalebox{0.7}{\includegraphics{zew_make.epsi}}
      \caption{The three topologies for $\mathscr{O}(\alpha_W^3)$ $Zqq$ production.}\label{fig:zew}
    \end{center}
  \end{minipage}
\end{figure}

\begin{figure}[ht]
  \begin{minipage}[t]{0.5\linewidth}
    \begin{center}
      \scalebox{0.58}{\includegraphics{zqcd1_make.epsi}}
      \caption{$\mathscr{O}(\alpha_s^2\alpha_W)$ $Zqq$ production.}\label{fig:zqcd}
    \end{center}
  \end{minipage}
  \hspace{1mm}
  \begin{minipage}[b]{0.5\linewidth}
    \begin{minipage}[t]{\linewidth}
      \begin{center}
        \scalebox{0.5}{\includegraphics{hbk_make.epsi}}
        \caption{$\mathscr{O}(\alpha_s^4)$ backgrounds to $qq\rightarrow qq(H,Z),\ (H,Z)\rightarrow b\bar{b}$.}\label{fig:hbk}
      \end{center}
      \vspace{6mm}
    \end{minipage}
    \begin{minipage}[t]{\linewidth}
      \begin{center}
        \scalebox{0.5}{\includegraphics
          {screen_make.epsi}}
        \caption{Screening of QCD dijet + $b\bar{b}$ production via gluon exchange.}\label{fig:screen}
      \end{center}
    \end{minipage}
  \end{minipage}
\end{figure}
  
\section{Parton Level Signal and Background Rates}
\begin{figure}
\centering
\includegraphics[angle=-90,width=75mm]
{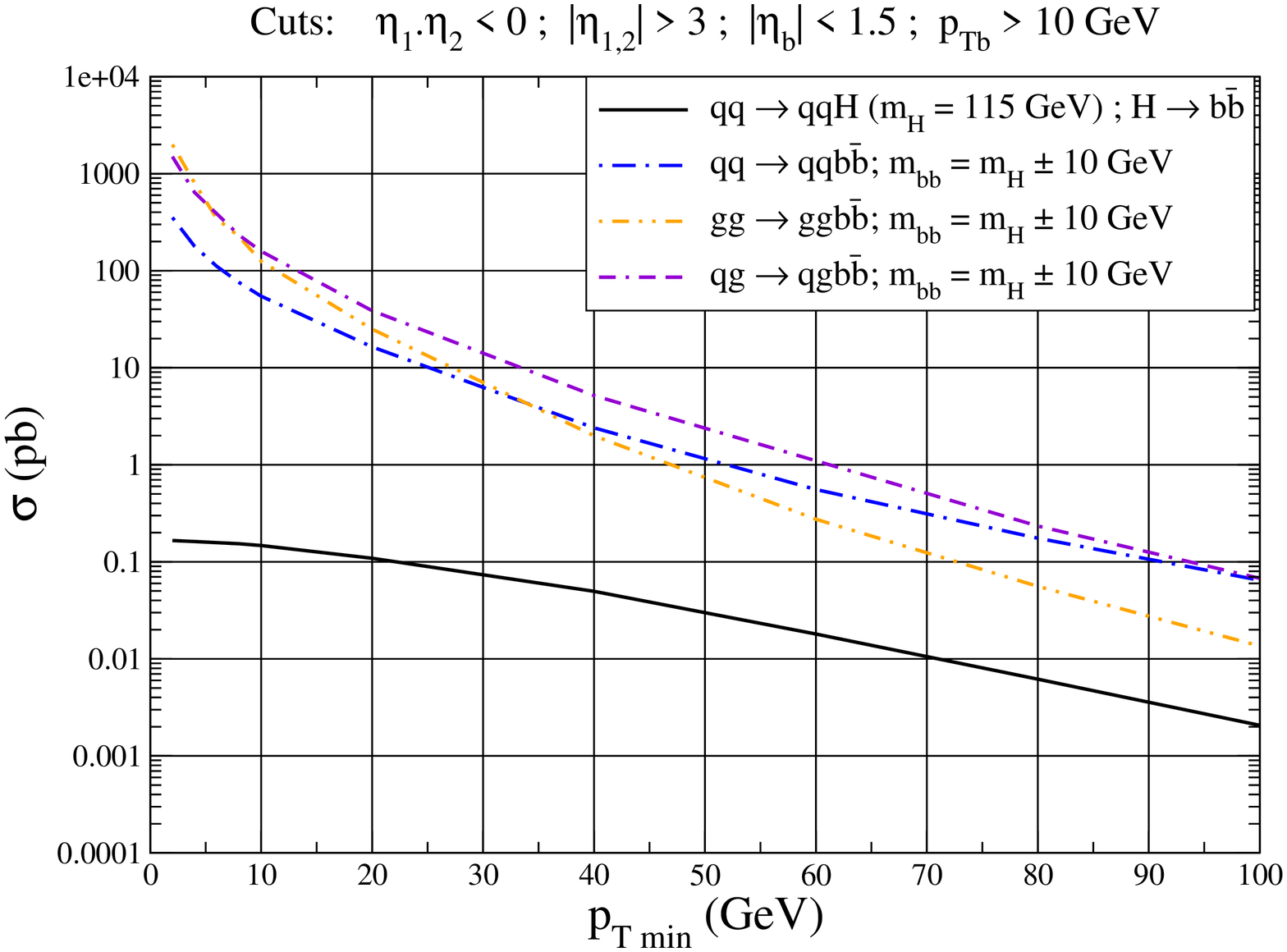}
\hspace{1mm}
\includegraphics[angle=-90,width=75mm]
{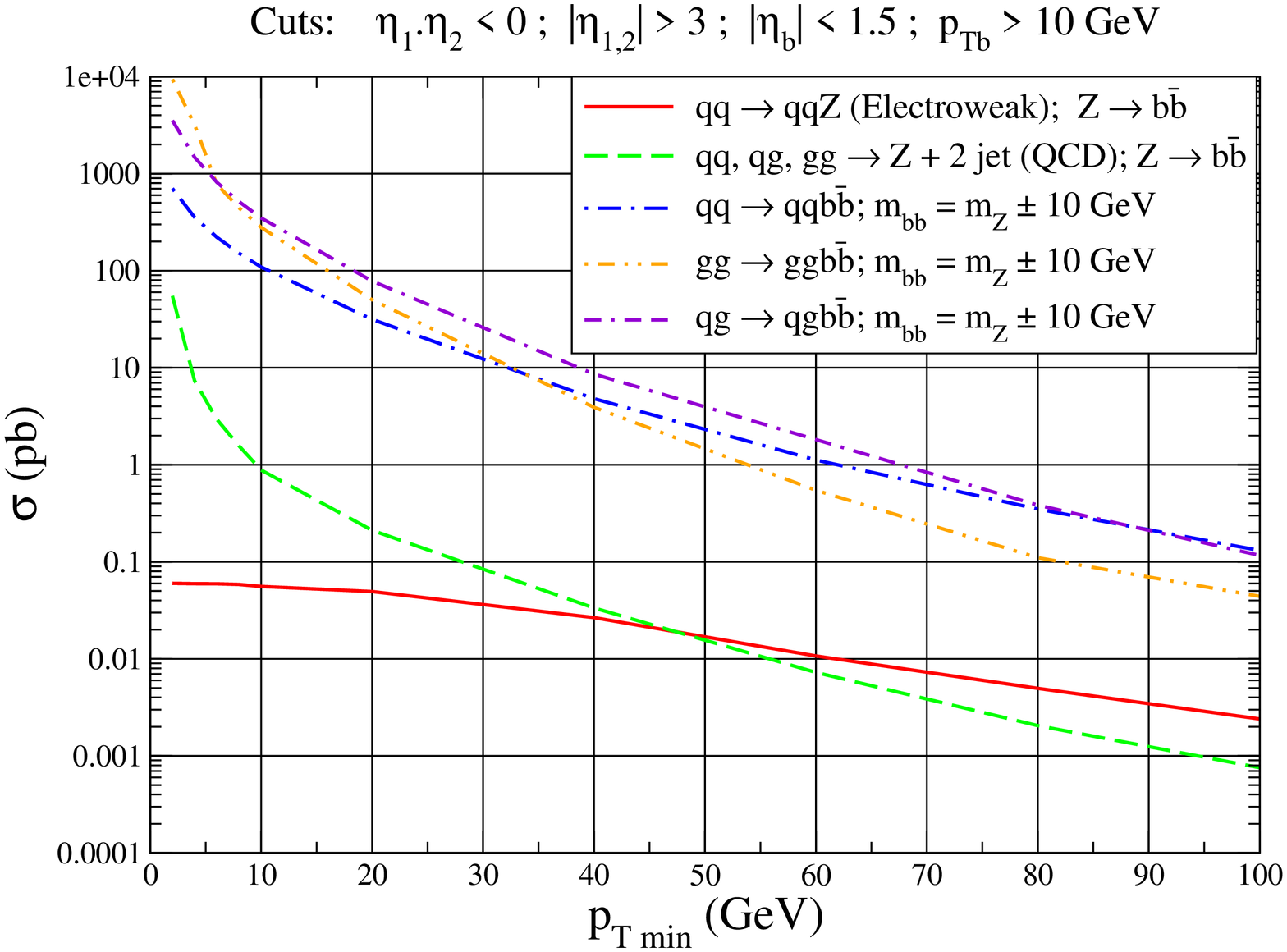}
\caption{Parton level cross sections at $\sqrt{s} = 14$ TeV for $H$ (left) and $Z$ (right)  production processes and backgrounds after application of cuts.}
\label{fig:cut_sig_Z}
\end{figure}

At the parton level, we define a set of selection cuts to pick out the
`natural' features of our two signals. Table~\ref{tab:loss} shows the effect
of applying these cuts in sequence to the Higgs signal. $p_{Tj}$ and $\eta_j$
are the transverse momentum and rapidity of the forward jet partons.  Of
course, the depletion of the backgrounds is much greater. The rates are
shown in Figure~\ref{fig:cut_sig_Z}.

\begin{table}[ht]
\caption{Loss of $qq\rightarrow qqH$ cross section at $\sqrt{s} = 14$~TeV with $M_{H}=115$~GeV in applying selection cuts and the $b \bar b$ branching ratio.}\label{tab:loss}
\vspace{0.4cm}
\begin{center}
\begin{tabular}{|c|c|c|}
\hline
Cut Imposed & Cross Section for $qq\rightarrow qqH$ at $p_{Tj} > 40$GeV &
 \% of Initial Cross Section \\
\hline\hline
  & $4.86$pb & 100\% \\
\hline
Br($H\rightarrow b\bar{b}$) & 3.49pb & 71.9\% \\ \hline
$\eta_{1}\cdot\eta_{2} < 0$ & 2.47pb & 50.8\% \\ \hline
$\Delta \eta_j > 6$ & 0.495pb & 10.2\% \\ \hline
$|\eta_{j}| > 3$ & 0.0990pb & 2.04\% \\ \hline
$|\eta_{b}| < 1.5$ & 0.0465pb & 0.957\% \\ \hline
$p_{Tb} > 10$~GeV & 0.0463pb & 0.953\% \\ \hline
\end{tabular}
\end{center}
\end{table}

\section{Gap Survival Probability}
\subsection{Parton Level}
The QCD-induced $b\bar{b}$ background is still large after cuts, exceeding by
two orders of magnitude the $Z/H$ cross sections. Further suppression can be
achieved by requiring a {\it completely clean} gap, i.e.  without any soft
hadrons. The only way to create a gap in a QCD induced event is to screen the
colour flow across the gap by an additional gluon; that is, to consider
graphs of the type shown in Fig.~\ref{fig:screen}. Using a leading
logarithmic approximation to this loop integral, one obtains the probability
to screen out the octet colour flow
\begin{equation}
\label{eqn:prob} P_{a}=C_{a}\left(\int^{\ptmin}_{Q_0}\alpha_{S}(Q_{T}^{2})\frac{dQ_{T}^{2}}{Q_{T}^2}\exp\left\{-\frac{N_{c}\Delta\eta}{2\pi}
\int_{Q_{T}}^{\ptmin}\alpha_{S}(Q^{\prime2})\frac{dQ^{\prime2}}{Q^{\prime2}}\right\}\right)^{2}
=C_{a}\left(\frac{2\pi}{N_{c}\Delta\eta}\right)^{2}. 
\end{equation}
Numerically, this suppresses the backgrounds by a factor of ten.\par
Another point we must take into account is the fact that now the
$b\bar{b}$-pair may be produced in a colour singlet state only, and the
ordinary $gg\rightarrow b\bar{b}$ hard subprocess cross section (which
includes both colour singlet and octet contributions) should be replaced by
the pure colour singlet cross section. Numerically, this
suppresses the background by a further factor of ten. Our arguments are
further explained in a recent paper~\cite{Khoze:2002fa}.
\subsection{Hadron Level}
We now require there to be no hadrons in the gaps at hadron level, i.e. we
must take into account soft interactions of spectator partons. 
Using the formalism of Ref.~\cite{19} we obtain for our kinematics
\begin{equation}
\label{eqn:surfacts}
\hat{S}_{Z}^{2}=0.31;\qquad\hat{S}_{H}^{2}=0.31;\qquad\hat{S}_{QCD b\bar{b}}^{2}=0.27. 
\end{equation}

\section{Results}

\begin{wrapfigure}[17]{r}[0mm]{70mm}
  \epsfig{file=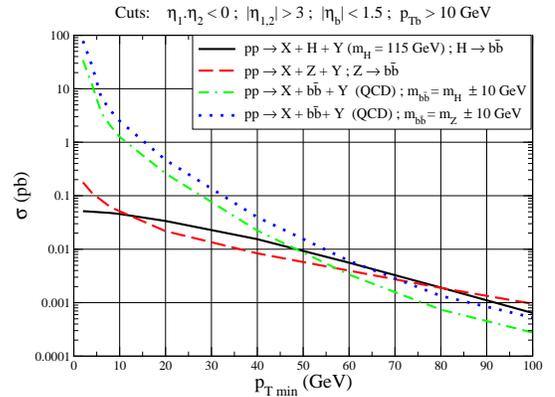, width=61mm,angle=-90}
  \caption{Hadron-level cross sections at $\sqrt{s} = 14$ TeV for inclusive
    Higgs and $Z$ production with subsequent decay to $b\bar{b}$ and their
    respective QCD $b\bar{b}$ backgrounds. These are plotted as a function of
    the minimum $p_{T}$ of the forward jets.}
  \label{fig:final_plot}

\end{wrapfigure}
Figure~\ref{fig:final_plot} summarises our results, which are based on the
parton level cross sections after application of cuts
(Fig.~\ref{fig:cut_sig_Z}). The QCD-induced cross sections (both the QCD $Z$
production of Fig.~\ref{fig:zqcd}d and the direct QCD $b\bar{b}$ production
of Fig.~\ref{fig:hbk}) are then multiplied by the probability to screen out
the colour octet contribution for the relevant initial state
(Eqn.~\ref{eqn:prob}). To take into account the fact that the $b\bar{b}$ pair
in the background processes can only be produced in the colour singlet state
the ordinary $gg\rightarrow b\bar{b}$ cross section is replaced by the pure
singlet cross section.  Finally both the signals and backgrounds are
multiplied by the relevant soft survival probability of
Eqn.~\ref{eqn:surfacts}.
\section*{Acknowledgements}
I have been fortunate to undertake this work with Valery Khoze, Misha Ryskin
and James Stirling.

\end{document}